\documentclass[12pt,showkeys]{article}

\usepackage{epsfig}

\begin{document}

\baselineskip=17pt plus 0.2pt minus 0.1pt

\makeatletter
\@addtoreset{equation}{section}
\renewcommand{\theequation}{\thesection.\arabic{equation}}
\def\a'{\alpha'}
\def\s{\sigma}
\def\t{\tau}
\def\tension{\frac{1}{4\pi\alpha'}}
\def\sumk{\sum^{\infty}_{k=1}}
\def\ust{u_*}
\def\ae{a_{\rm eff}}
\def\phieff{\phi_{\rm eff}}
\def\calM{{\cal M}}
\def\calO{{\cal O}}
\def\calV{{\cal V}}
\def\calD{{\cal D}}
\def\p{{\partial}}
\def\nn{{\nonumber}}
\newcommand{\bea}{\begin{eqnarray}}
\newcommand{\eea}{\end{eqnarray}}
\begin{titlepage}
\title{
\hfill\parbox{4cm}
{\normalsize {\tt hep-th/0406193v4}}\\
\vspace{1cm}
{\bf Recoiling D-branes
 } 
}
\author{
Shin {\sc Nakamura}
\thanks{{\tt E-mail: nakamura@nbi.dk}}
\\[15pt]
{\it The Niels Bohr Institute}\\
{\it Blegdamsvej 17, DK-2100 Copenhagen \O, Denmark}\\
\\[10pt]
}
\date{\normalsize June, 2004}
\maketitle
\thispagestyle{empty}

\begin{abstract}
\normalsize
\noindent
We propose a new method to describe a recoiling D-brane 
that is elastically
scattered by closed strings in the non-relativistic region.
We utilize the low-energy effective field theory on the worldvolume 
of the D-brane, and the velocity of the D-brane is 
described by the time derivative of the expectation values 
of the massless scalar fields on the worldvolume.
The effects of the closed strings are represented 
by a source term for the massless fields in this method.
The momentum conservation condition between the closed strings 
and the D-brane is derived up to the relative sign
of the momentum of the D-brane.

\end{abstract}

\end{titlepage}

\section{Introduction}

Studies of interactions between D-branes and closed strings are 
quite important from various points of view.
The interactions 
play a crucial role in finding non-trivial
relationships between open strings and closed strings such as
AdS/CFT \cite{AdS/CFT} and recently-proposed open-closed 
dualities \cite{Sen,GIR,BR}, for example. 
Studies of the interactions are also important to 
analyze the dynamics of the systems of multiple D-branes. 

However, almost all analyses of the interactions have been 
carried out by considering static D-branes, and it is a 
long-standing problem to describe the back reaction of the D-branes.
A D-brane in the worldsheet description is 
just a boundary of the worldsheet with Dirichlet boundary 
condition, and the D-brane is treated as an infinitely heavy
classical source of closed strings; 
the position, or the collective coordinate, of the D-brane
is fixed at a point in the target space in the Dirichlet direction.

There are several attempts to describe the back reaction of
D-branes. 
Some of them are based on the conformal field theory on
the worldsheet \cite{HK,FPR-1,FPR-2,PT}.
In Ref. \cite{HK}, the disk amplitude for the
scattering of closed string states from a D-particle is computed 
in the bosonic string theory in which the
collective coordinate of the D-particle is quantized.
The current conservation condition of the D-particle is
obtained by demanding the conformal invariance of the
amplitude in that work.
Momentum conservation condition
between the closed strings and the D-particle is described by
using the zero-mode integral in the path integral of the
trajectory of the D-particle.

Attempts to obtain the momentum conservation condition
between closed strings and a D-brane 
from the viewpoint of conformal invariance
can be found in Refs.
\cite{FPR-1,FPR-2,PT}. An annulus amplitude for the 
scattering of the closed strings from the D-brane is considered
and a variant of the Fischler-Susskind mechanism is proposed
there; the infrared (IR) divergence in the open string channel 
of the annulus amplitude is canceled by adding an appropriate 
operator to the boundary of the worldsheet.
The momentum conservation condition between the closed strings
and the D-particle is explicitly obtained 
in Refs. \cite{FPR-2,PT} 
by demanding the conformal invariance of the total amplitude.
In Ref. \cite{PT}, the IR divergence
is canceled by adding
a logarithmic operator \cite{Gur} that represents 
the recoil of the D-particle \cite{KM,KMW}.\footnote{
For other related studies on D-brane recoil with
the logarithmic operators, see
Refs. \cite{EMN-1,EMN-2,MJ-1,MS-1,EMEW,MS-2,MJ-2},
 for example.}
However, it is also pointed out in Ref. \cite{PT}
that the divergence does not 
exist in the case of D$p$-brane with $p>1$. 
In Ref. \cite{FPR-2}, 
the energy conservation condition
is also obtained as well as the momentum 
conservation condition in the case of D-particle, explicitly.
However, there is still room to clarify
how to define the initial momentum of the D-particle there.
Further investigation along the ideas of Refs. \cite{FPR-1,FPR-2,PT}
is still important for deeper understanding of recoil
of D-branes.
Some applications of D-brane recoil to other topics, and
related works are found in Refs. 
\cite{EKMNW,EMN-3,CEMN,CM-0,EMNV,CM,EMN-4,MW,LM,GM-2,GM,EMW}.

A target-space theory that handles second-quantized closed strings
with dynamical D-branes may provide us a description of
back reaction of D-branes.
Some arguments on back reaction and recoil of D-branes along this
approach is found in Ref. \cite{AKM}.

In the present work, we propose an alternative method to 
describe the scattering process between a D-brane and closed strings 
together with the back reaction of the D-brane 
in the bosonic string theory.
The impact of the closed strings in this method is represented 
by a source term of the low-energy effective worldvolume theory 
of the D-brane, and
the initial and the final velocity of the D-brane are 
described by the time derivative of the expectation values 
of the massless scalar fields of the worldvolume theory.
We utilize the following approximations to justify our
approach:
\begin{enumerate}
  \item Field theory limit, namely $k^{2}\alpha'\ll  1$ where
  $k$ is the typical momentum of the open strings on the
  D-brane.
  \item Elastic limit, namely the momenta of the closed strings
   are small enough and no massive open-string mode is excited
   on the D-brane.
   We also assume that the closed strings do not lose their
   total momentum in the worldvolume directions of the D-brane, 
   and no internal
   field on the D-brane gets momentum from the closed strings. 
  \item Non-relativistic limit, namely the velocity of the 
  D-brane is very small.
  \item Tree level approximation in the string theory, namely
   the string coupling is very small.
\end{enumerate}
The above conditions 3 and 4 means that the tension of
the D-brane is very large. This is also consistent with the
condition 1. 
 
A nontrivial problem is how to represent the source term
of the worldvolume theory in terms of the quantities of the
closed strings.
The basic idea is as follows.
We consider a scattering process between the D-brane and 
the closed strings that creates $n$ massless scalar particles 
on the worldvolume of the D-brane. 
We calculate the probability of the creation of the $n$ 
massless scalar particles 
in the two different frameworks: one of them is
the worldvolume theory of the D-brane with the source term and
the other is the perturbative string theory.
By comparing the two results, we obtain the relationship
between the source term and the momenta of the closed
strings, and we obtain 
the momentum conservation condition between the closed strings 
and the D-brane up to the relative sign of the momentum of
the D-brane.
One of the distinction between the present work and those
in Refs. \cite{FPR-1,FPR-2,PT} is that
all the diagrams we consider in the string theory are
disk diagrams and we need not annulus diagrams.

The organization of this article is as follows.
We consider bosonic strings, and we start by considering
a recoiling D-particle for simplicity.
In section 2, we consider the worldvolume theory of the
D-particle and review some basic facts necessary for the later
discussions.
We calculate the amplitude of the creation of the $n$ massless 
particles explicitly.
In section 3, we consider the scattering process between
the D-particle and the closed strings and calculate
the amplitude of the process that creates $n$ massless
open string modes on the D-particle in the framework of the
string theory.
In section 4, we compare the results obtained in section 2 with
those in section 3.
We show that the absolute value of the total momentum transfer 
from the closed strings
is exactly equal to the absolute value of the change of 
the D-particle's momentum,
within the above approximations.
We also comment on the distribution of the probability
of the $n$ massless particle creation. We see that the expectation
value of the total energy of the created particles gives
the kinetic energy of the D-particle, correctly.
In section 5, we generalize the results obtained for the
D-particle to the case of higher dimensional D-branes.
We compactify the worldvolume of the D-brane
to make its mass finite, and we obtain the momentum conservation 
condition between the closed strings and the D-brane up to the 
relative sign of the momentum of the D-brane, again.
We provide conclusion, several open problems and discussions
in the last section.

\section{Effective field theory on D-particle}
\label{field}

\subsection{Effective action}

We begin with the Dirac-Born-Infelt (DBI) action of a D-particle.
It is just an ordinary action for a point particle:
\begin{eqnarray}
S_{\mbox{\scriptsize DBI}}=-\tau \int dt \sqrt{1-\dot{X}^{2}},
\end{eqnarray}
where $\tau$ is the tension of the D-particle, 
and $\dot{X}^{\mu}$ denotes the time derivative of 
the space coordinate $X^{\mu}$ of the D-particle.
We consider a {\em non-relativistic} D-particle in this work.
Then $\dot{X}^{2} \ll 1$ and the DBI action is written as
\begin{eqnarray}
S_{\mbox{\scriptsize DBI}}
=\tau \int dt \left\{\frac{1}{2}(\partial_{t} X)^{2} \right\},
\end{eqnarray}
where we have dropped the constant term and the $O(\dot{X}^{4})$
contributions.

Next, we consider a scattering process between the D-particle 
and closed strings. 
We attempt to include the effects of the closed strings 
in the above action.
A simple conjecture is that a source term may
effectively represent the impact of the closed strings.
Therefore, we consider the following effective action
\begin{eqnarray}
S=\tau \int dt 
\left\{
\frac{1}{2}(\partial_{t} X)^{2}-\frac{1}{2}m^{2}X^{2}+J(t)\cdot X(t)
\right\},
\label{effective}
\end{eqnarray}
and let us examine how this action represents the
nature of the recoiling D-particle.
Here, $m$ is an IR cut-off that should be sent 
to zero finally, and we assume that $J(t)$ includes 
relevant effects of the closed strings.

\subsection{Expectation value of $\dot{X}^{\mu}$}

Let us calculate the time dependence of the expectation
value of $X^{\mu}$ at the classical level.
We choose the target space coordinate so that the
initial momentum of the D-particle is equal to zero
and its final momentum is in the $x^{25}$ direction.
We assume that the source term is switched on only
for the duration 
$t_{\mbox{\scriptsize i}}\le t \le t_{\mbox{\scriptsize f}}$.
The expectation value of the field $X^{25}$ is given by
\begin{eqnarray}
X^{25}(t)=X^{25}_{0}(t)+\int^{\infty}_{-\infty}dt'
G_{\mbox{\scriptsize ret}}(t-t')J^{25}(t'),
\end{eqnarray}
where $G_{\mbox{\scriptsize ret}}(t-t')$ is 
the retarded Green's function which satisfies
\begin{eqnarray}
(\partial^{2}_{t}+m^{2})G_{\mbox{\scriptsize ret}}(t-t')
=\theta(t-t')\delta(t-t'),
\end{eqnarray}
and $X^{25}_{0}(t)$ satisfies the equation of motion without 
the source term,
\begin{eqnarray}
(\partial^{2}_{t}+m^{2})X^{25}_{0}(t)=0.
\end{eqnarray}
$\dot{X}^{25}_{0}$ is the initial velocity of the
D-particle and $\dot{X}^{25}_{0}=0$, in fact.
Substituting the explicit form of the retarded Green's function,
\begin{eqnarray}
X^{25}(t)=X^{25}_{0}(t)+\int^{\infty}_{-\infty}dt'
\int^{\infty}_{-\infty}
\frac{dq}{2\pi}\frac{e^{iq(t-t')}}{-(q-i\epsilon)^{2}+m^{2}}
J^{25}(t'),
\end{eqnarray}
where we should take the limit $\epsilon \to +0$.
Then,
\begin{eqnarray}
\dot{X}^{25}=\dot{X}^{25}_{0}
+\int^{\infty}_{-\infty}
\frac{dq}{2\pi i}\frac{q\: e^{iqt}}{(q-i\epsilon+m)(q-i\epsilon-m)}
\tilde{J}^{25}(q),
\end{eqnarray}
where $\tilde{J}^{25}(q)$ is the source term in the momentum
space.\footnote{
Our definition of the Fourier transformation is
\begin{eqnarray}
F(t)&=&\int^{\infty}_{-\infty}\frac{dq}{2\pi}e^{iqt}\tilde{F}(q),\\
\tilde{F}(q)&=&\int^{\infty}_{-\infty}dt \:e^{-iqt}F(t).
\end{eqnarray}
}
Note that $\dot{X}^{25}=\dot{X}^{25}_{0}=0$
for $t\le t_{\mbox{\scriptsize i}}$.

For $t \ge t_{\mbox{\scriptsize f}}$,
\begin{eqnarray}
\dot{X}^{25}
&\to&
\frac{1}{2}e^{imt}\tilde{J}^{25}(m)
+\frac{1}{2}e^{-imt}\tilde{J}^{25}(-m)
\:\:\:\:\:\:(\epsilon \to +0)
\nonumber \\
&\to&
\tilde{J}^{25}(0)\:\:\:\:\:\:(m \to 0).
\label{X-dot}
\end{eqnarray}
The limit $\epsilon \to +0$ should be taken before we take the
limit $m \to 0$.
Note that $\tilde{J}^{25}(0)$ is real.
$t_{\mbox{\scriptsize f}}-t_{\mbox{\scriptsize i}}$ 
should be regarded as the duration
of that the closed strings affect the D-particle in the
scattering process. We assume that
the effects of the closed strings disappear at the
infinitely far future and we define the final velocity
of the D-particle $\dot{X}^{25}_{\mbox{\scriptsize f}}$ 
in this region ($t \ge t_{\mbox{\scriptsize f}}$).
In the same way, we assume that the effects of the closed
strings disappears at the infinitely past and
the initial velocity of the D-particle 
$\dot{X}^{25}_{\mbox{\scriptsize i}}$
should be defined in the region of 
$t \le t_{\mbox{\scriptsize i}}$.

Then, the change of the velocity of the D-particle 
is given by 
$\dot{X}^{25}_{\mbox{\scriptsize f}}=\tilde{J}^{25}(0)$, 
and the problem we should consider is how to rewrite 
$\tilde{J}^{25}(0)$ in terms of the quantities of 
the closed strings.
We will come back to this problem in section 
\ref{conservation-section}.

\subsection{$n$ particle creation amplitude}

We calculate the $n$ particle creation amplitude with the
source term in order to compare it with the corresponding
amplitude in the string theory.
We treat the source term as a perturbation.

The action without the source term is
\begin{eqnarray}
S_{0}=\tau \int dt 
\left\{
\frac{1}{2}(\partial_{t} X)^{2}-\frac{1}{2}m^{2}X^{2}
\right\}.
\end{eqnarray}
We consider only $X^{25}$ and we define a dimension-less field 
$\phi(t)$ as $X^{25}=\sqrt{2\pi \alpha'}\phi$.
Then the Lagrangian and the canonical momentum are
\begin{eqnarray}
L_{0}&=&2\pi \alpha' \tau
\left\{
\frac{1}{2}(\partial_{t} \phi)^{2}-\frac{1}{2}m^{2}\phi^{2}
\right\}, \\
{\cal P}&=&2\pi \alpha' \tau \dot{\phi} \:.
\end{eqnarray}
We define the creation operator $a$ and the annihilation operator 
$a^{\dagger}$ as
\begin{eqnarray}
\phi&=&\frac{1}{\sqrt{2h}}(a+a^{\dagger}),\\
{\cal P}&=&-i\sqrt{\frac{h}{2}}(a-a^{\dagger}),
\end{eqnarray}
where $h\equiv 2\pi \alpha' \tau m$.
The Hamiltonian is given by
\begin{eqnarray}
H_{0}=m\left(a^{\dagger}a+\frac{1}{2}\right).
\end{eqnarray}
Including the source term, the Hamiltonian is
\begin{eqnarray}
H&=&H_{0}+V,\\
V&=&
-j(t)(a+a^{\dagger}),
\end{eqnarray}
where we have defined $j(t)$ as
\begin{eqnarray}
j(t)\equiv \sqrt{\frac{\tau}{2m}} J^{25}(t).
\label{j-J}
\end{eqnarray}
We define the $n$ particle state as
$|n\rangle=\frac{1}{\sqrt{n!}}(a^{\dagger})^{n}|0\rangle$.

By using the above notations,
the $n$ particle creation amplitude 
${\cal A}^{(n)}_{\mbox{\scriptsize field}}$ is given by
\begin{eqnarray}
{\cal A}^{(n)}_{\mbox{\scriptsize field}}
&\equiv&
\lim_{t_{0} \to -\infty, t \to \infty}
\langle n|e^{-iH(t-t_{0})}|0\rangle
\nonumber \\
&=&
\frac{1}{\sqrt{n!}}(i)^{n}\tilde{j}(0)^{n}
e^{-\frac{1}{2}|\tilde{j}(0)|^{2}}
e^{i\theta},
\label{A-field}
\end{eqnarray}
at the limit $m \to 0$.
$\theta$ is a real number defined in (\ref{theta}).
Note that 
$|{\cal A}^{(n)}_{\mbox{\scriptsize field}}|^{2}
=\frac{1}{n!}\lambda^{n} e^{-\lambda}$ 
gives a Poisson distribution
where $\lambda=|\tilde{j}(0)|^{2}$, and the final state
is a coherent state.
The detailed calculation to obtain (\ref{A-field}) is
reviewed in appendix \ref{n-parti-amp}.

\section{Calculations in string theory}
\label{string}

\begin{figure}[h]
  \begin{center}
   \epsfig{figure=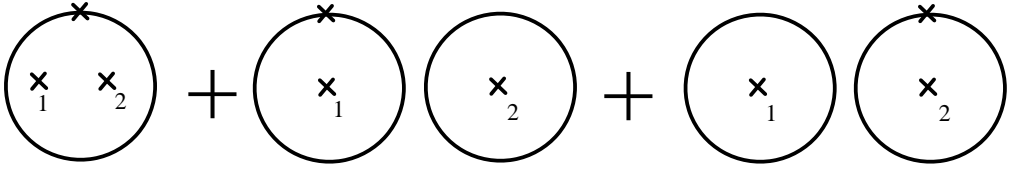}
  \end{center}
  \caption{The diagrams of the amplitude of one open-string 
  creation from two closed strings.}
  \label{fig:diagrams}
\end{figure}

In this section,
we calculate the amplitude of the creation of the $n$
massless scalar open strings on the D-particle from
$N$ closed strings, at the tree level.
We choose the spacetime coordinate so that the D-particle
recoils in the $x^{25}$ direction as we did in section
\ref{field}, and the massless open strings we consider here
correspond to the zero mode of 
the D-particle in the $x^{25}$ direction.\footnote{
The massless open strings discussed in the following
should always be understood as the zero modes of the
D-brane in the $x^{25}$ direction.
}
We also choose the spacetime coordinate so that the initial
velocity of the D-particle is zero.
The amplitude consists of several disk diagrams as shown 
in Fig. \ref{fig:diagrams}, for example.
We calculate a single disk amplitude to begin with, and 
we consider the full diagrams by using the results on 
the single disk amplitude.

\subsection{Single disk amplitude}

We calculate a disk amplitude $A^{(N,n)}$ of $N$ closed strings 
and $n$ massless scalar open strings on the D-particle.
We consider a unit disk ${\cal M}$ with Neumann boundary
condition $\partial_{r}X^{0}|_{r=1}=0$ and Dirichlet boundary
conditions $X^{\nu}|_{r=1}=0$ for $\nu \neq 0$, where
$z=r e^{i\phi}$ is the complex coordinate on the unit disk.
We explicitly calculate for the case that all the closed strings 
are closed-string tachyons, for simplicity, however the
results should be easily generalized to the cases of arbitrary
closed strings.
The open strings cannot carry momenta in the Dirichlet directions 
and thus their momenta in the space directions are zero.
The energy of the massless open 
string is also zero due to the on-shell condition.
Therefore, the massless open string labeled by $l$ is represented
by the vertex operator 
$i\zeta_{(l)}\cdot \partial_{r} X(e^{i\phi_{l}})$
where $\zeta_{(l)}^{\nu}$ is the polarization vector
which satisfies
$\zeta_{(l)}^{25}=1$ and $\zeta_{(l)}^{\nu}=0$ for $\nu \neq 25$.

\subsubsection{$N\ge 2$ case}

The amplitude is given by
\begin{eqnarray}
A^{(N,n)}&=&
i g_{\mbox{\scriptsize c}}^{N}g_{\mbox{\scriptsize o}}^{n}
\int_{0}^{1}dr_{2}
\left(\prod_{i=3}^{N} \int d^{2}z_{i} \right)
\left(\prod_{l=N+1}^{N+n} \int_{0}^{2\pi} d\phi_{l} \right)
\nonumber \\
&&
\left \langle
c(0)\tilde{c}(0)e^{ik_{(1)}\cdot X(0)}
c^{\phi}(r_{2})e^{ik_{(2)}\cdot X(r_{2})}
\prod_{i=3}^{N} e^{ik_{(i)}\cdot X(z_{i})}
\prod_{l=N+1}^{N+n} i\zeta_{(l)}^{25}\partial_{r}X^{25}(e^{i \phi_{l}})
\right \rangle_{{\cal M}}
\nonumber \\
&=&
i g_{\mbox{\scriptsize c}}^{N}g_{\mbox{\scriptsize o}}^{n}
C\delta(\sum_{i}k_{(i)}^{0})
\nonumber \\
&& \times
\left.
\int_{0}^{1}dr_{2}
\left(\prod_{i=3}^{N} \int d^{2}z_{i} \right)
Z_{\mbox{\scriptsize gh}}(r_{2})
\prod_{i<j\le N}|z_{i}-z_{j}|^{\alpha'k_{(i)}^{\mu}{k_{(j)}}_{\mu}}
\prod_{i,j\le N}
|1-z_{i}\bar{z_{j}}|^{a(i,j)}
\right|_{z_{1}=0,z_{2}=r_{2}}
\nonumber \\
&& \times
\left(\prod_{l=N+1}^{N+n} \int_{0}^{2\pi} d\phi_{l} \right)
\prod_{l=N+1}^{N+n}
\prod_{i=1}^{N}
\exp\{-k_{(i)}^{25}\zeta_{(l)}^{25}
\langle X^{25}(z_{i}) \partial_{r}X^{25}(e^{i\phi_{l}})\rangle  \}
\nonumber \\
&& \times
\prod_{l,m=N+1,\: l< m}^{N+n}
\exp\{-\zeta_{(m)}^{25}\zeta_{(l)}^{25}
\langle \partial_{r}X^{25}(e^{i\phi_{m}}) 
\partial_{r}X^{25}(e^{i\phi_{l}})\rangle  \}
,
\label{Amplitude}
\end{eqnarray}
where only the terms that are linear with respect to every
$\zeta_{(l)}^{25}$ ($l=N+1,\cdots,N+n$) should be taken
and the other terms should be discarded in (\ref{Amplitude}).
$C$ is a constant,
$g_{\mbox{\scriptsize o}}$ is the coupling constant for 
the massless open strings and
$g_{\mbox{\scriptsize c}}$ is the closed-string coupling constant.
$c$, $\tilde{c}$ and $c^{\phi}$ is the ghost fields of
holomorphic, anti-holomorphic and the $\phi$-component
respectively.
We have fixed the positions of the
closed string vertices as $z_{1}=0$ and $z_{2}=r_{2}$,
where $z_{j}=r_{j}e^{i \phi_{j}}$ is the position of the
$j$-th vertex operator.\footnote{
$\phi_{2}=0$ is fixed and $r_{2}$ can still move in the region 
$0\le r_{2} \le 1$.}
The worldsheet coordinates are assigned as follows:
$z_{j}$'s with $1\le j \le N$ are for the closed strings
and $z_{j}$'s with $N+1\le j \le N+n$ are for the open strings.
We have also defined $a(i,j)$ as
\begin{eqnarray}
a(i,j)\equiv  \frac{\alpha'}{2} 
  \left(k_{(i)}^{0}{k_{(j)}^{0}}
  -\sum_{\nu\neq 0}k_{(i)}^{\nu}{k_{(j)}^{\nu}}\right)
,
\end{eqnarray}
and $Z_{\mbox{\scriptsize gh}}(r_{2})$ as
\begin{eqnarray}
Z_{\mbox{\scriptsize gh}}(r_{2})
\equiv
\langle c(0)\tilde{c}(0) c^{\phi}(r_{2})
\rangle_{{\cal M}}.
\end{eqnarray}
We can easily find that $Z_{\mbox{\scriptsize gh}}(r_{2})$ 
is a constant for the present case.
Note that 
\begin{eqnarray}
\int d\phi_{m} \int d\phi_{l} 
\zeta_{(m)}^{25}\zeta_{(l)}^{25}
\langle \partial_{r}X^{25}(e^{i\phi_{m}}) 
\partial_{r}X^{25}(e^{i\phi_{l}})\rangle
=0
\label{vanish-int}
\end{eqnarray}
due to (\ref{vanish}), and the last line of (\ref{Amplitude})
becomes $1$.
We can calculate the contribution of the third line 
of (\ref{Amplitude}) by using
\begin{eqnarray}
&&\int_{0}^{2\pi} d\phi_{l}
\sum_{i=1}^{N}
k_{(i)}^{25}\zeta_{(l)}^{25}
\langle X^{25}(z_{i}) \partial_{r}X^{25}(e^{i\phi_{l}})\rangle
\nonumber \\
&=&
-\int_{0}^{2\pi} d\phi_{l}
\sum_{i=1}^{N}
k_{(i)}^{25}\zeta_{(l)}^{25}
\frac{\alpha'}{2}
\left\{
\frac{w}{w-z_{i}}-\frac{w}{w-1/\bar{z_{i}}}
+
\frac{\bar{w}}{\bar{w}-\bar{z_{i}}}-\frac{\bar{w}}{\bar{w}-1/z_{i}}
\right\}
\nonumber \\
&=&
-2\pi \alpha' \sum_{i=1}^{N}k_{(i)}^{25},
\end{eqnarray}
where $w=e^{i\phi_{l}}$, $|z_{i}|\le 1$ and we have substituted 
$\zeta_{(l)}^{25}=1$.
See appendix \ref{notations} for the details.
Therefore the amplitude is given by
\begin{eqnarray}
A^{(N,n)}&=&
i g_{\mbox{\scriptsize c}}^{N}g_{\mbox{\scriptsize o}}^{n}
C\delta(\sum_{i}k_{(i)}^{0})Z_{\mbox{\scriptsize gh}}
\nonumber \\
&& \times
\left.
\int_{0}^{1}dr_{2}
\left(\prod_{i=3}^{N} \int d^{2}z_{i} \right)
\prod_{i<j\le N}|z_{i}-z_{j}|^{\alpha'k_{(i)}^{\mu}{k_{(j)}}_{\mu}}
\prod_{i,j\le N}
|1-z_{i}\bar{z_{j}}|^{a(i,j)}
\right|_{z_{1}=0,z_{2}=r_{2}}
\nonumber \\
&& \times
\prod_{l=N+1}^{N+n}
\left(
2\pi \alpha' \sum_{i=1}^{N}k_{(i)}^{25}
\right)
\nonumber \\
&=&
\left(
2\pi \alpha' g_{\mbox{\scriptsize o}}\sum_{i=1}^{N}k_{(i)}^{25}
\right)^{n}
\times
A^{(N,0)}
\nonumber \\
&=&
\left(
\frac{1}{\sqrt{\tau}}\sum_{i=1}^{N}k_{(i)}^{25}
\right)^{n}
\times
A^{(N,0)},
\end{eqnarray}
where we have used the relationship 
$\tau (2\pi \alpha' g_{\mbox{\scriptsize o}})^{2}=1$ 
in the last line.
$A^{(N,0)}$ is the disk amplitude of $N$ closed-string
tachyons without open strings.
Note that the momentum conservation condition among the
strings appears due to the zero-mode integral.
We do not have the momentum conservation condition among the
strings in the Dirichlet directions due to lack of the
zero-mode integral and thus 
$\sum_{i=1}^{N}k_{(i)}^{25}\neq 0$ is allowed.

\subsubsection{$N=1$ case}

In this case, we should fix both the position of the 
closed-string vertex and that of one of the open-string
vertices.\footnote{
Alternative method of the calculation is to leave the positions
of the open-string vertices unfixed and divide the amplitude
by the volume of the rotational symmetry as mentioned bellow
(\ref{identif}).
}
We will fix the position of the open-string vertex labeled
by $l=2$ as $z_{2}=e^{i\phi_{2}}$, where $\phi_{2}$ is a 
fixed value. The position of the closed-string vertex
will be fixed at the center of the unit disk, again.

The amplitude is given by
\begin{eqnarray}
A^{(1,n)}&=&
i g_{c}g_{o}^{n}
C\delta(k_{(1)}^{0})
Z_{\mbox{\scriptsize gh}}
\frac{1}{2\pi}
\prod_{l=2}^{1+n}
\left(
2\pi \alpha' k_{(1)}^{25}
\right)
\nonumber \\
&=&
\left(
\frac{1}{\sqrt{\tau}}k_{(1)}^{25}
\right)^{n}
\times
A^{(1,0)},
\label{fact-amp-1}
\end{eqnarray}
where we have identified $A^{(1,0)}$ as
\begin{eqnarray}
A^{(1,0)}=
i g_{c}
C\delta(k_{(1)}^{0})
Z_{\mbox{\scriptsize gh}}
\frac{1}{2\pi}.
\label{identif}
\end{eqnarray}
The factor $1/(2\pi)$ in the first line of (\ref{fact-amp-1})
appears because of the absence of the $\phi_{2}$ integral.
The identification (\ref{identif}) can be justified as follows.
In the calculation of $A^{(1,0)}$, we fix the position
of the closed-string vertex at the center of the disk.
Although the degrees of freedom of the modular transformation
on the disk have not fixed completely, the residual symmetry is
just a rotational symmetry the volume of which is $2\pi$.
Then let us fix the rotational degree of freedom by hand.
More precisely, let us mark at the point $z=e^{i\phi}$, for
example, and fix the value of the $\phi$-coordinate of
the marked point by hand.
This corresponds to divide the amplitude by $2\pi$.
Now we have fixed the position of the closed string vertex
and the $\phi$-coordinate of the marked point on the disk.
Thus the Faddeev-Popov determinant is the same as the
case we have more than one vertex; it is
$Z_{\mbox{\scriptsize gh}}$.
As a result, we obtain (\ref{identif}).

\subsubsection{General results}

After all, we obtain 
\begin{eqnarray}
A^{(N,n)}
=
\left(
\frac{1}{\sqrt{\tau}}\sum_{i=1}^{N}k_{(i)}^{25}
\right)^{n}
\times
A^{(N,0)},
\label{fact-amp-pre}
\end{eqnarray}
for all positive integers $N$.
All the closed-string vertices we have used in the 
calculations are closed-string tachyon vertices, 
however we can easily find that
(\ref{fact-amp-pre}) holds for arbitrary
closed strings if we regard $A^{(N,n)}$ as the disk 
amplitude of the corresponding $N$ closed strings
and the $n$ massless open strings.

\subsection{Total amplitude}

Now we can calculate the total amplitude of the creation of
the $n$ massless scalar open strings by using the results
obtained in the previous subsection.
We define the total amplitude as ${\cal A}^{(N,n)}$.
For simplicity, we demonstrate the calculation of the amplitude 
in the case of $N=2$ and $n=1$ that is given schematically 
in Fig. \ref{fig:diagrams}.
We can easily calculate the total amplitude by using
(\ref{fact-amp-pre}) as
\begin{eqnarray}
{\cal A}^{(2,1)}
&=&
A^{(2,1)}
+A^{(1,1)}_{(1)}
A^{(1,0)}_{(2)}
+A^{(1,0)}_{(1)}
A^{(1,1)}_{(2)}
\nonumber \\
&=&
\frac{\sum_{i=1}^{2}k^{25}_{(i)}}{\sqrt{\tau}}
A^{(2,0)}
+
\frac{k^{25}_{(1)}}{\sqrt{\tau}}
A^{(1,0)}_{(1)}
A^{(1,0)}_{(2)}
+
A^{(1,0)}_{(1)}
\frac{k^{25}_{(2)}}{\sqrt{\tau}}
A^{(1,0)}_{(2)}
\nonumber \\
&=&
\frac{\sum_{i=1}^{2}k^{25}_{(i)}}{\sqrt{\tau}}
\left\{
A^{(2,0)}
+
A^{(1,0)}_{(1)}
A^{(1,0)}_{(2)}
\right\}
\nonumber \\
&=&
\frac{\sum_{i=1}^{2}k^{25}_{(i)}}{\sqrt{\tau}}
{\cal A}^{(2,0)},
\end{eqnarray}
where ${\cal A}^{(2,0)}$ is the total amplitude for no open-string
creation and $A^{(1,n)}_{(i)}$ is the
disk amplitude with $n$ massless open-string vertices and
$i$-th closed-string vertex.
The above calculation can be easily extended into the case
of $N>2$, and we obtain
\begin{eqnarray}
\frac{{\cal A}^{(N,1)}}{{\cal A}^{(N,0)}}
=
\frac{\sum_{i}k^{25}_{(i)}}{\sqrt{\tau}},
\label{fact-amp}
\end{eqnarray}
in general.
Furthermore, we can show that
\begin{eqnarray}
\frac{{\cal A}^{(N,n)}}{{\cal A}^{(N,0)}}
=
\left (
\frac{\sum_{i}k^{25}_{(i)}}{\sqrt{\tau}}
\right)^{n}.
\label{string-poisson}
\end{eqnarray}
The derivation of (\ref{fact-amp}) and (\ref{string-poisson}) 
is given in appendix \ref{poisson}. 

\section{Momentum conservation}
\label{conservation-section}

\subsection{The source term from the string amplitude and
the momentum conservation}
\label{normalization}

In order to rewrite $\tilde{J}^{25}(0)$
in terms of the quantities of the closed strings,
we compare the amplitude obtained in section \ref{field}
with that in section \ref{string}.
Before comparing them explicitly, we should notice the
following facts:
\begin{enumerate}
  \item \underline{Normalization of the states}\\
If we have $n$ same particles in the final state,
we should divide the cross section by $n!$.
In other words,
we should multiply $1/\sqrt{n!}$
to the string amplitude in order to compare
it with the amplitude in the worldvolume theory where the factor
$1/\sqrt{n!}$ is already included in the definition of the
$n$-particle state.
  \item \underline{Dimension of the spacetime}\\
In the worldvolume theory, the massless particle exists in the
one-dimensional worldvolume and the calculation in section
\ref{field} is based on non-relativistic quantum mechanics.
On the other hand, the strings considered in section \ref{string}
exist in 26-dimensional spacetime, and the center-of-mass
coordinates of the massless open strings
are confined in the one-dimensional subspace of it.
Because of the above difference, we also need the following
consideration to obtain the correct normalization.

Let us consider the probability $P(n)$ of that $n$ open-string
particles of
mass $m$ whose momenta lie in a small region 
$d^{25}\vec{k}_{(1)}\cdots d^{25}\vec{k}_{(n)}$
are created from the closed strings.\footnote{
We have introduced the same IR cut-off as that of the worldvolume 
theory here; now the mass of the open-string particles is $m$.
On the other hand, we have calculated the string amplitudes
for the case of $m=0$ in order to maintain the conformal invariance
of the worldsheet.
This discrepancy disappears because we take the limit
$m \to 0$ eventually.
}
$P(n)$ is given as
\begin{eqnarray}
P(n)&=&\frac{d^{25}\vec{k}_{(1)}}{(2\pi)^{25}}
\frac{1}{2E_{\vec{k}_{(1)}}}
\cdots 
\frac{d^{25}\vec{k}_{(n)}}{(2\pi)^{25}}\frac{1}{2E_{\vec{k}_{(n)}}}
\:
\left|\frac{1}{\sqrt{n!}}{\cal A}^{(N,n)}\right|^{2}
\:
(2\pi)^{25}\delta(\vec{k}_{(1)})
\cdots
(2\pi)^{25}\delta(\vec{k}_{(n)})
\nonumber \\
&=&
\left(
\prod_{i=1}^{n}
\frac{1}{2E_{\vec{k}_{(i)}}}\right)
\left|\frac{1}{\sqrt{n!}}{\cal A}^{(N,n)}\right|^{2}
\nonumber \\
&=&
\left(\frac{1}{2m} \right)^{n} 
\left|\frac{1}{\sqrt{n!}}{\cal A}^{(N,n)}\right|^{2}.
\label{P(n)}
\end{eqnarray}
We have the delta functions in the first line because
the open strings do not have momenta in the Dirichlet
directions.
Therefore, we should also multiply the factor 
$\left(\frac{1}{2m} \right)^{n/2}$ to the amplitude
in the sting theory to compare it with the amplitude in
the worldvolume theory. 
Note that the factor $(\frac{1}{2m})^{n}$ in (\ref{P(n)})
is necessary to make $P(n)$ dimensionless.

  \item \underline{Difference of the Hilbert spaces}\\
The Hilbert space of string theory in section \ref{string}
includes the closed-string sector and the massive open-string
sector, too.
On the other hand, the Hilbert space of the effective field
theory on the D-particle considered in section
\ref{field} includes only massless open-string states.
Due to the above difference of the Hilbert spaces,
the absolute normalization of the amplitudes
cannot be compared directly with each other.
\end{enumerate}

With the above things in mind, we propose the following
identification between the amplitudes
in the effective field theory and those
in the string theory:
\begin{eqnarray}
\left |
\frac{{\cal A}^{(1)}{\mbox{\scriptsize field}}}
{{\cal A}^{(0)}{\mbox{\scriptsize field}}}
\right |^{2}
=
\frac{P(1)}{P(0)}
&=&
\left |
\frac{\frac{1}{\sqrt{2m}}\frac{1}{\sqrt{1!}}{\cal A}^{(N,1)}}
{\frac{1}{\sqrt{0!}}{\cal A}^{(N,0)}}
\right |^{2}
\nonumber \\
&=&
\frac{1}{2m}
\left |
\frac{{\cal A}^{(N,1)}}{{\cal A}^{(N,0)}}
\right |^{2}
.
\label{amp-compare}
\end{eqnarray}
The physical meaning of the right-hand side
is the following.
The probability of the creation of the one open string
through the scattering between the D-particle and the
closed strings should be proportional to
$\frac{1}{2m}|{\cal A}^{(N,1)}|^{2}$ and that of no creation
of the massless open string should be proportional to
$|{\cal A}^{(N,0)}|^{2}$.
We can cancel the ambiguity of the normalization of the
probability which arises due to the problem 3 above,
by considering the ratio of the probability.

Let us substitute the above results into (\ref{amp-compare}).
From Eqs. (\ref{X-dot}), (\ref{j-J}), (\ref{A-field}) 
and (\ref{fact-amp}), we obtain
\begin{eqnarray}
\frac{{\cal A}^{(1)}{\mbox{\scriptsize field}}}
{{\cal A}^{(0)}{\mbox{\scriptsize field}}}
&=&
i \: \tilde{j}(0), 
\\
\frac{{\cal A}^{(N,1)}}{{\cal A}^{(N,0)}}
&=&
\frac{1}{\sqrt{\tau}}
\sum_{i=1}^{N}k_{(i)}^{25}, 
\label{str-amp-ratio}
\\
\tau \dot{X}^{25}_{\mbox{\scriptsize f}}&=&
\sqrt{2m\tau}\: \tilde{j}(0).
\end{eqnarray}
Then we obtain the following relationship from (\ref{amp-compare}):
\begin{eqnarray}
\left | \tau \dot{X}^{25}_{\mbox{\scriptsize f}} \right |
=\left | \sum_{i=1}^{N}k_{(i)}^{25} \right |.
\label{conservation}
\end{eqnarray}
This is the momentum conservation condition between the D-particle
and the closed strings up to the relative sign.
The result (\ref{conservation}) is independent of the
IR cut-off and of the types of the closed strings.

\subsection{Poisson distribution from string theory and the
kinetic energy of the D-particle}

We have considered only $P(1)/P(0)$ in the previous subsection.
Let us consider $P(n)/P(0)$ for general $n$ and see what
happens.
In the framework of the effective field theory,
$P(n)$ is given from (\ref{A-field}) as
\begin{eqnarray}
P(n)&=&\frac{1}{n!}\lambda^{n}e^{-\lambda},
\\
\lambda&=&|\tilde{j}(0)|^{2},
\end{eqnarray}
and $P(n)/P(0)=\frac{1}{n!}\lambda^{n}$.
The above gives a Poisson distribution.
In the framework of the string theory, $P(n)/P(0)$ is given as
\begin{eqnarray}
\frac{P(n)}{P(0)}=
\frac{1}{n!}
\left(
\frac{1}{\sqrt{2m}}\frac{\sum_{i}k^{25}_{(i)}}{\sqrt{\tau}}
\right)^{2n},
\label{st-poisson}
\end{eqnarray}
from (\ref{string-poisson}) and the discussions 
in section \ref{normalization}.
Thus the identification 
\begin{eqnarray}
|\tilde{j}(0)|
=\frac{1}{\sqrt{2m}}\frac{\sum_{i}k^{25}_{(i)}}{\sqrt{\tau}},
\end{eqnarray}
that is made in (\ref{amp-compare}) is still valid for $n\ge 2$.
In other words, (\ref{st-poisson}) indicates that we can
also derive the Poisson distribution for the massless particles
from the string theory.
The above observation strongly suggests that the effective
action (\ref{effective}) with the source term captures
the phenomena of the scattering between the D-particle 
and the closed strings well.

It is interesting to see that the total energy of the
massless scalar particles on the worldvolume of the D-particle
gives the correct kinetic energy of the D-particle.
The expectation value of the total energy of the
scalar particles after the scattering is given
by using (\ref{st-poisson}) as 
\begin{eqnarray}
m \frac{\sum_{n=0}^{\infty}nP(n)}{\sum_{n=0}^{\infty}P(n)}
=
\frac{\left(\sum_{i}k^{25}_{(i)}\right)^{2}}{2\tau},
\label{kinetic-e}
\end{eqnarray}
that is exactly the expected value of the kinetic energy of
the D-particle after the scattering.
This quantity is also independent of the IR cut-off.

\section{Generalization to the higher dimensional D-branes}

We generalize the results in the previous sections to the
cases of higher dimensional D-branes.
We consider a D$p$-brane which extends in the 
$x^{0}, \cdots, x^{p}$ directions.
We choose the spacetime coordinate so that the initial momentum
of the D-brane is
zero and the final momentum of it is in the $x^{25}$ direction. 
We consider an elastic scattering between the D$p$-brane and
closed strings; we assume that the momenta of the closed strings
parallel to the D$p$-brane are conserved within the closed strings,
and the total momentum coming from the closed strings is exactly
perpendicular to the D$p$-brane.
In this case, the 
D$p$-brane does not get any internal energy from the closed strings.

\subsection{Results from the worldvolume theory}
\label{h-field}

In this section, we consider the following effective scalar 
field theory on the D$p$-brane:
\begin{eqnarray}
S=\tau_{p} \int d^{p+1}x 
\left\{
\frac{1}{2}\partial_{a} X_{\rho}\partial^{a} X^{\rho}
-\frac{1}{2}m^{2}X^{2}+J(x)\cdot X(x)
\right\},
\label{effective-h}
\end{eqnarray}
where $m$ is an IR cut-off and $\tau_{p}$ is the tension
of the D$p$-brane. 
$\rho$ runs from $p+1$ to $25$.
We have chosen the worldvolume coordinate 
$x^{a}$ so that it is equal to the zero mode of $X^{a}$,
where $a$ runs from $0$ to $p$. 
We have $U(1)$ gauge field on the worldvolume and the gauge
field may contribute to the recoiling process of the D-brane,
too. However, we assume that all the contributions of the
gauge field are implicitly included in the source term.

We also assume that the source term is switched on only for
the duration 
$t_{\mbox{\scriptsize i}} \le x^{0} \le t_{\mbox{\scriptsize f}}$.
We compactify the $x^{1}, \cdots, x^{p}$ directions on a torus
of radius $R$ to make the mass of the D$p$-brane finite. 
Let us consider momentum conservation in the $x^{25}$ direction
and we consider only the field $X^{25}$.
Standard calculations in the field theory lead us to the
following results:
\begin{eqnarray}
P(n)_{\mbox{\scriptsize D}p}
&=&
\frac{1}{n!}\lambda_{p}^{n}
e^{-\lambda_{p}},
\label{A^2-field-h}
\\
&&\lambda_{p}
\equiv
\frac{\tau_{p}}{V_{p}}
\sum_{\vec{n}}
\frac{1}{2E_{\vec{q}}}
|\tilde{J}^{25}(E_{\vec{q}},\vec{q})|^{2},
\label{lambda-p-q}
\\
\dot{X}^{25}_{\mbox{\scriptsize f}}
&=&
\frac{1}{2 V_{p}}
\sum_{\vec{n}}
\left\{\tilde{J}^{25}(E_{\vec{q}},\vec{q})
e^{iE_{\vec{q}} x^{0}}e^{-i\vec{q} \cdot \vec{x}}
+\tilde{J}^{25}(-E_{\vec{q}},-\vec{q})
e^{-iE_{\vec{q}} x^{0}}e^{i\vec{q} \cdot \vec{x}}
\right\},
\label{X-dot-h-q}
\end{eqnarray}
where 
$P(n)_{\mbox{\scriptsize D}p}$
is the probability of that the source creates
$n$ massless scalar particles that correspond to
the zero mode of the D$p$-brane in the $x^{25}$ direction,
and $\dot{X}^{25}_{\mbox{\scriptsize f}}$ 
is the time derivative of the expectation value of 
$X^{25}$ at $x^{0} \ge t_{\mbox{\scriptsize f}}$.
We regard $\dot{X}^{25}_{\mbox{\scriptsize f}}$ 
as the final velocity of the D-brane in the $x^{25}$ 
direction as we did in the case of D-particle.
The integrals over the momenta in the uncompactified theory
have been replaced with the summation over the Kaluza-Klein momenta 
$\vec{q}=(\frac{n_{1}}{R},\cdots, \frac{n_{p}}{R})$ which are
labeled by $\vec{n}\equiv (n_{1},\cdots, n_{p})$.
Note that the volume of the D$p$-brane $V_{p}=(2\pi R)^{p}$ 
has appeared in order to get the correct measure of the
summation.
We have defined $E_{\vec{q}}$ and $\vec{x}$ as
$E_{\vec{q}}=\sqrt{|\vec{q}|^{2}+m^{2}}$ and
$\vec{x}=(x^{1}, \cdots, x^{p})$.
$\tilde{J}^{25}(E_{\vec{q}},\vec{q})$ is the source term
in the momentum space that is given by
\begin{eqnarray}
\tilde{J}^{25}(E_{\vec{q}},\vec{q})
=
\int^{\infty}_{-\infty}dx^{0}
\int^{2\pi R}_{0} \!\!\!\!\! dx^{1}
\cdots
\int^{2\pi R}_{0} \!\!\!\!\! dx^{p}
e^{-iE_{\vec{q}}x^{0}+i\vec{q} \cdot \vec{x}}
J(x^{0},\vec{x}) .
\end{eqnarray}

The probability that the source creates the
one massless scalar particle of momentum $\vec{k}$
is given by
\begin{eqnarray}
P(1)_{\mbox{\scriptsize D}p}
&=&
\lambda_{p,\vec{k}}
e^{-\lambda_{p}},
\\
\lambda_{p,\vec{k}}
&\equiv&
\frac{\tau_{p}}{V_{p}}
\frac{1}{2E_{\vec{k}}}
|\tilde{J}^{25}(E_{\vec{q}},\vec{k})|^{2}.
\label{lambda-p-k}
\end{eqnarray}
We are considering the elastic case in which
the D$p$-brane does not get any internal energy from
the closed strings. 
Then $\tilde{J}^{25}(E_{\vec{q}},\vec{k})=0$ for $\vec{k}\neq 0$,
because only the massless particle with $\vec{k}=0$ can be
created in this process.
Therefore Eqs. (\ref{lambda-p-q}) and (\ref{X-dot-h-q}) become
\begin{eqnarray}
\lambda_{p}
&=&
\frac{\tau_{p}}{V_{p}}
\frac{1}{2m}
|\tilde{J}^{25}(0,0)|^{2},
\label{lambda-h}
\\
\dot{X}^{25}_{\mbox{\scriptsize f}}
&=&
\frac{1}{V_{p}}
\tilde{J}^{25}(0,0).
\label{X-dot-h}
\end{eqnarray}
in the present case.

\subsection{Comparison with the amplitude of string theory }

In the case that the massless scalar open string does not carry
the momentum, we can easily see that the calculations 
in the string theory gives a similar result as that of the D-particle
case:
\begin{eqnarray}
\left |
\frac{{\cal A}^{(N,1)}_{\mbox{\scriptsize D}p}}
{{\cal A}^{(N,0)}_{\mbox{\scriptsize D}p}}
\right |^{2}
&=&
\frac{1}{\tau_{p}}
\left( \sum_{i=1}^{N}k_{(i)}^{25} \right)^{2},
\label{A1/A0-h}
\end{eqnarray}
where $A^{(N,n)}_{\mbox{\scriptsize D}p}$ is the total tree-level
amplitude of the creation of the $n$ massless scalar open 
strings on the D$p$-brane from $N$ closed strings.\footnote{
These massless open strings are those corresponding to
the zero mode of the D$p$-brane in the $x^{25}$ direction.
}
Next, we generalize the relationship (\ref{amp-compare}) to
the present case.
Let us consider the probability of that one scalar particle of mass 
$m$ whose momenta lies in a small region $d^{25}\vec{k}$ 
is created from the closed strings.\footnote{
$\vec{k}$ is the momentum of the massless particle as that in
section \ref{h-field}, however it is defined as a 25-dimensional
vector in this subsection.} 
The probability written
in terms of the string amplitude up to the normalization is
\begin{eqnarray}
\frac{d^{25}\vec{k}}{(2\pi)^{25}}\frac{1}{2E_{\vec{k}}}
|{\cal A}^{(N,1)}_{\mbox{\scriptsize D}p}|^{2}
\prod_{\rho=p+1}^{25}\Bigl( 2\pi\delta(k^{\rho}) \Bigr)
=
\frac{d^{p}\vec{k}}{(2\pi)^{p}}
\frac{1}{2E_{\vec{k}}}|{\cal A}^{(N,1)}_{\mbox{\scriptsize D}p}|^{2},
\label{prob}
\end{eqnarray}
where we have considered uncompactified spacetime here.
The delta functions come from the fact that the scalar
particle do not have the momentum perpendicular to the D$p$-brane.
In the present case, the spacetime is compactified and $1/V_{p}$
appears when we convert the momentum integrals into the summation
over the Kaluza-Klein momenta. Then the probability (\ref{prob}) 
should be written as
\begin{eqnarray}
\frac{1}{V_{p}}
\frac{1}{2m}|{\cal A}^{(N,1)}_{\mbox{\scriptsize D}p}|^{2},
\end{eqnarray}
where we have substituted $\vec{k}=0$ as we did bellow
(\ref{lambda-p-k}).
The same probability in terms of the field theory
is given by $P(1)_{\mbox{\scriptsize D}p}$, 
and then the relationship corresponding to (\ref{amp-compare})
for the present case is
\begin{eqnarray}
\frac{P(1)_{\mbox{\scriptsize D}p}}{P(0)_{\mbox{\scriptsize D}p}}
=
\frac{1}{2m V_{p}}
\left |
\frac{A^{(N,1)}_{\mbox{\scriptsize D}p}}
{A^{(N,0)}_{\mbox{\scriptsize D}p}}
\right |^{2}
.
\label{relation-h}
\end{eqnarray}
From Eqs. (\ref{A^2-field-h}), (\ref{lambda-h}), 
(\ref{X-dot-h}), (\ref{A1/A0-h}) and (\ref{relation-h}),
we obtain
\begin{eqnarray}
\left|
\tau_{p}V_{p}
\dot{X}^{25}_{\mbox{\scriptsize f}}
\right|
=
\left|\sum_{i=1}^{N}k_{(i)}^{25}\right|,
\end{eqnarray}
which is the correct momentum conservation condition in the
direction perpendicular to the D$p$-brane up to the relative sign.

We can also show that the Poisson distribution of the probability 
of the $n$ massless particle creation obtained from the
effective field theory is consistent with the calculations of 
the creation probabilities in the string theory.
The expectation value of the total energy of the created
massless particles gives the final kinetic
energy of the D$p$-brane correctly.

\section{Conclusion and discussions}

We developed a new method to describe an elastic scattering
between a D-brane and closed strings in the non-relativistic
region. The momentum conservation condition between the D-brane 
and the closed strings was obtained up to the relative sign
of the D-brane's momentum.

An interesting extension of the present work may be 
investigation of non-elastic scatterings 
in which the closed strings lose their total momentum
in the worldvolume directions.
In this case, the massless modes on 
the D-brane get non-zero momentum and non-zero energy
from the closed strings.
How to obtain the energy conservation condition
between the D-brane and the closed strings,
that have not been discussed in this paper,
is also an important problem which we should solve in the future.
If we choose a particular spacetime coordinate so that 
$\dot{X}^{25}_{\mbox{\scriptsize i}}=
-\dot{X}^{25}_{\mbox{\scriptsize f}}$,
energy conservation holds through the scattering process.
This suggests that the Lorentz covariant generalization
of the present work may solve the problem of the energy
conservation.
Generalization to superstring theory is interesting, too.

In the present work, the momentum conservation condition
between the D-brane and the closed strings was {\em derived}
up to the relative sign and 
a natural question is how to fix the relative sign.
However, we can take a different standpoint in the construction
of more generalized formalism to solve the above open problems;
the momentum conservation condition with correct 
relative sign can be utilized as a prerequisite.
For example, the exact momentum conservation condition
for a D-particle,
\begin{eqnarray}
\tau \dot{X}^{25}_{\mbox{\scriptsize f}}
=\sum_{i=1}^{N}k_{(i)}^{25},
\label{abs-conservation}
\end{eqnarray}
fixes the relative phase between
the amplitudes in the field theory and those in the string theory
as 
\begin{eqnarray}
\frac{{\cal A}^{(1)}{\mbox{\scriptsize field}}}
{{\cal A}^{(0)}{\mbox{\scriptsize field}}}
=i
\frac{1}{\sqrt{2m}}
\frac{{\cal A}^{(N,1)}}{{\cal A}^{(N,0)}},
\label{abs-amp-comp}
\end{eqnarray}
in our notation.
We can also obtain a similar condition between the amplitudes
for a D$p$-brane.
The new relationship between the amplitudes like
(\ref{abs-amp-comp}) can be a starting point of
solving the open problems.
We hope that the present work can be a step to the
deeper understanding of the interactions between
D-branes and closed strings.

\vspace{0.5cm}
\noindent
{\large\bf Acknowledgments}

The author would like to thank John F. Wheater for the
collaboration at the initial stage of this project.
The author also thanks the members of the group of 
theoretical particle physics and cosmology at Niels Bohr Institute 
and NORDITA for valuable discussions and comments.
The author wish to thank the organizers of RTN workshop
``The quantum structure of spacetime and the geometric 
nature of fundamental interactions''
held at Copenhagen on 15-20 September 2003 where
the present work has been initiated.
This work was supported in part 
by Nishina Memorial Foundation.

\appendix

\section{Derivation of (\ref{A-field})}
\label{n-parti-amp}

We review the derivation of (\ref{A-field}) in this appendix.
In the interaction representation,
\begin{eqnarray}
|\Psi(t)\rangle_{I}=U(t,t_{0})|\Psi(t_{0})\rangle_{I},
\end{eqnarray}
where
\begin{eqnarray}
U(t,t_{0})&=&\mbox{T} 
\exp\left\{-i \int^{t}_{t_{0}}dt' V_{I}(t')\right\}
\nonumber \\
&=&
\exp\left\{i a^{\dagger}\int^{t}_{t_{0}}dt' j(t')e^{imt'}\right\}
\exp\left\{i a\int^{t}_{t_{0}}dt' j(t')e^{-imt'}\right\}
\nonumber \\
&&\times \exp\left\{-\int^{t}_{t_{0}}dt_{1}
\int^{t_{1}}_{t_{0}}dt_{2}\: \theta(t_{1}-t_{2})e^{im(t_{1}-t_{2})
}j(t_{1})j(t_{2})\right\}.
\end{eqnarray}
The time evolution of the state in the Schr\"{o}dinger
representation is
\begin{eqnarray}
|\Psi(t)\rangle_{S}=e^{-iH_{0}(t-t_{0})}
U(t,t_{0})|\Psi(t_{0})\rangle_{S}.
\end{eqnarray}

The $n$ particle state is defined as
$|n\rangle=\frac{1}{\sqrt{n!}}(a^{\dagger})^{n}|0\rangle$
and
\begin{eqnarray}
\langle n|e^{-iH(t-t_{0})}|0\rangle
&=&\langle n|e^{-iH_{0}(t-t_{0})}U(t,t_{0})|0\rangle
\nonumber \\
&=&e^{-im(n+\frac{1}{2})(t-t_{0})}
\langle n|e^{ia^{\dagger}\alpha}|0\rangle
\nonumber \\
&&\times
\exp\left\{-\int^{t}_{t_{0}}dt_{1}
\int^{t_{1}}_{t_{0}}dt_{2}\: \theta(t_{1}-t_{2})e^{im(t_{1}-t_{2})
}j(t_{1})j(t_{2})\right\},
\end{eqnarray}
where
\begin{eqnarray}
\alpha=\int^{t_{1}}_{t_{0}}dt' j(t')e^{imt'}.
\end{eqnarray}
The following formulae are useful:
\begin{eqnarray}
\langle n|e^{ia^{\dagger}\alpha}|0\rangle
&=&\frac{1}{\sqrt{n!}}(i)^{n}(\alpha)^{n}.
\\
\theta(t_{1}-t_{2})
&=&\frac{1}{2\pi i}
\int^{\infty}_{-\infty}\frac{dq}{q-i\epsilon}e^{iq(t_{1}-t_{2})}
\nonumber \\
&=&\frac{1}{2\pi i}
\int^{\infty}_{-\infty}dq
\left\{\mbox{P}\frac{1}{q}+i\pi\delta(q)\right\}
e^{iq(t_{1}-t_{2})}.
\end{eqnarray}
Therefore
\begin{eqnarray}
&&-\int^{t}_{t_{0}}dt_{1}
\int^{t_{1}}_{t_{0}}dt_{2}\: \theta(t_{1}-t_{2})e^{im(t_{1}-t_{2})
}j(t_{1})j(t_{2})
\nonumber \\
&\to&
-\frac{1}{2\pi i}P
\int^{\infty}_{-\infty}\frac{dq}{q}
\tilde{j}(q-m)\tilde{j}(-q+m)
-\frac{1}{2}\tilde{j}(-m)\tilde{j}(m)
\nonumber \\
&&\:\:\:\:(t_{0} \to -\infty, t \to \infty)
\nonumber \\
&\to&
i\theta-\frac{1}{2}|\tilde{j}(0)|^{2}
\:\:\:\:(m \to 0),
\end{eqnarray}
where
\begin{eqnarray}
\theta \equiv 
\frac{1}{2\pi}\mbox{P}
\int^{\infty}_{-\infty}\frac{dq}{q}
\tilde{j}(q)\tilde{j}(-q)
\label{theta}
\end{eqnarray}
is a real number.
Thus $n$ particle creation amplitude 
${\cal A}^{(n)}_{\mbox{\scriptsize field}}$ is given as
\begin{eqnarray}
{\cal A}^{(n)}_{\mbox{\scriptsize field}}
&\equiv&
\lim_{t_{0} \to -\infty, t \to \infty}
\langle n|e^{-iH(t-t_{0})}|0\rangle
\nonumber \\
&=&
\frac{1}{\sqrt{n!}}(i)^{n}\tilde{j}(0)^{n}
e^{-\frac{1}{2}|\tilde{j}(0)|^{2}}
e^{i\theta},
\label{A-field-append}
\end{eqnarray}
at the limit $m \to 0$.

\section{Notations and useful formulae} 
\label{notations}

We use the following worldsheet action in this paper:
\begin{eqnarray}
S&=&\frac{1}{2\pi\alpha'}\int d^{2}z
\partial X^{\mu}\bar{\partial}X^{\nu}G_{\mu\nu},\\
& &G_{\mu\nu}
={\rm diag}(+,+,+,\cdots+).
\end{eqnarray}
The Green's function on the unit disk is given as
\begin{eqnarray}
\langle X^{\mu}(z)X^{\mu}(w) \rangle
=-\frac{\alpha'}{2}
\left\{
\ln|z-w|^{2}+\ln|1-z\bar{w}|^{2}
\right\}
\end{eqnarray}
if we have Neumann boundary condition in the $x^{\mu}$ direction,
and
\begin{eqnarray}
\langle X^{\mu}(z)X^{\mu}(w) \rangle
=-\frac{\alpha'}{2}
\left\{
\ln|z-w|^{2}-\ln|1-z\bar{w}|^{2}
\right\}
\end{eqnarray}
in the case of Dirichlet boundary condition instead.

We assume that we have Dirichlet boundary condition
in the $x^{25}$ direction here.
If the point $w$ is on the boundary of the unit disk,
\begin{eqnarray}
\langle X^{25}(z)\partial_{r}X^{25}(w) \rangle
&=&
-\frac{\alpha'}{2}
\left\{
\frac{w}{w-z}-\frac{w}{w-1/\bar{z}}
+
\frac{\bar{w}}{\bar{w}-\bar{z}}-\frac{\bar{w}}{\bar{w}-1/z}
\right\},
\end{eqnarray}
where 
$\partial_{r}X^{\mu}(w)
=(w\partial_{w}+\bar{w}\partial_{\bar{w}})X^{\mu}(w)$.
If the point $z$ is also on the boundary,
\begin{eqnarray}
\langle \partial_{r}X^{25}(z)\partial_{r}X^{25}(w) \rangle
=-\alpha'
\left\{
\frac{wz}{(w-z)^{2}}+
\frac{\bar{w}\bar{z}}{(\bar{w}-\bar{z})^{2}}
\right\}.
\label{cor-3}
\end{eqnarray}
Eq. (\ref{vanish-int}) can be shown by using (\ref{cor-3})
and
\begin{eqnarray}
\int_{0}^{2\pi} d\phi
\frac{e^{i\phi} e^{i\theta}}{(e^{i\phi}-e^{i\theta})^{2}}
&=&
e^{i\theta}
\left[
\frac{-1/i}{e^{i\phi}-e^{i\theta}}
\right]^{2\pi}_{0}
\nonumber \\
&=&0.
\label{vanish}
\end{eqnarray}

\section{Derivation of (\ref{fact-amp}) and (\ref{string-poisson}) }
\label{poisson}

We derive (\ref{fact-amp}) and (\ref{string-poisson}) 
in this appendix.
To begin with, we define $A(m_{1},m_{2},\cdots,m_{j};n )$
as the disk amplitude with $m_{1}$-th, $m_{2}$-th,
$\cdots$, and $m_{j}$-th closed-string vertices and the
$n$ massless scalar open-string vertices.
For example,
\begin{eqnarray}
A(1,3,4;n)=
\left(
\frac{k^{25}_{(1)}+k^{25}_{(3)}+k^{25}_{(4)}}{\sqrt{\tau}}
\right)^{n}
A(1,3,4;0),
\end{eqnarray}
by using (\ref{fact-amp-pre}).
${\cal A}^{(N,0)}$ is then written by the summation of all the
possible connected and disconnected diagrams:
\begin{eqnarray}
{\cal A}^{(N,0)}
&=&
A(1,2,\cdots,N;0)
\nonumber \\
&+&
A(1,2,\cdots,N-1;0)A(N;0)
+A(1,2,\cdots,N-2,N;0)A(N-1;0)
\nonumber \\
&&
+\cdots
+A(2,3,\cdots,N;0)A(1;0)
\nonumber \\
&+&
A(1,2,\cdots,N-2;0)\{ A(N-1,N;0)+A(N-1;0)A(N;0)\}
+\cdots
\nonumber \\
&+&\cdots
\nonumber \\
&+&
A(1)A(2) \cdots A(N).
\label{cal-A-N-0}
\end{eqnarray}
Next, let us calculate ${\cal A}^{(N,1)}$
by using a modification of (\ref{cal-A-N-0}).
In order to obtain ${\cal A}^{(N,1)}$, we should modify,
for example, $A(1,2,\cdots,N-2;0)A(N-1;0)A(N;0)$ in the right-hand
side of (\ref{cal-A-N-0}) as
\begin{eqnarray}
&&A(1,2,\cdots,N-2;0)A(N-1;0)A(N;0) 
\Longrightarrow 
\nonumber \\
&&\:\:\:\: A(1,2,\cdots,N-2;1)A(N-1;0)A(N;0)
+A(1,2,\cdots,N-2;0)A(N-1;1)A(N;0)
\nonumber \\
&&\:\:\:\: +A(1,2,\cdots,N-2;0)A(N-1;0)A(N;1)
\nonumber \\
&&\:\:\:\: =
\left\{
\frac{\sum_{i=1}^{N-2}k^{25}_{(i)}}{\sqrt{\tau}}
+\frac{k^{25}_{(N-1)}}{\sqrt{\tau}}
+\frac{k^{25}_{(N)}}{\sqrt{\tau}}
\right\}
A(1,2,\cdots,N-2;0)A(N-1;0)A(N;0)
\nonumber \\
&&\:\:\:\: =
\frac{\sum_{i=1}^{N}k^{25}_{(i)}}{\sqrt{\tau}}
A(1,2,\cdots,N-2;0)A(N-1;0)A(N;0),
\end{eqnarray}
and the extra factor 
$\frac{\sum_{i=1}^{N}k^{25}_{(i)}}{\sqrt{\tau}}$ 
appears in front of the original term.
In the same way, we can show that all the terms in the
right-hand side of (\ref{cal-A-N-0}) get the same extra factor
through the modification and we can conclude that
\begin{eqnarray}
{\cal A}^{(N,1)}
=\frac{\sum_{i=1}^{N}k^{25}_{(i)}}{\sqrt{\tau}}
{\cal A}^{(N,0)}.
\end{eqnarray}

${\cal A}^{(N,2)}$ is also calculated by a similar modification
of ${\cal A}^{(N,1)}$.
For example, ${\cal A}^{(N,1)}$ contains 
$A(1,2,\cdots,N-2;1)A(N-1;0)A(N;0)$ and this term should
be modified as
\begin{eqnarray}
&&A(1,2,\cdots,N-2;1)A(N-1;0)A(N;0) 
\Longrightarrow 
\nonumber \\
&&\:\:\:\: A(1,2,\cdots,N-2;2)A(N-1;0)A(N;0)
+A(1,2,\cdots,N-2;1)A(N-1;1)A(N;0)
\nonumber \\
&&\:\:\:\: +A(1,2,\cdots,N-2;1)A(N-1;0)A(N;1)
\nonumber \\
&&\:\:\:\: =
\left\{
\frac{\sum_{i=1}^{N-2}k^{25}_{(i)}}{\sqrt{\tau}}
+\frac{k^{25}_{(N-1)}}{\sqrt{\tau}}
+\frac{k^{25}_{(N)}}{\sqrt{\tau}}
\right\}
A(1,2,\cdots,N-2;1)A(N-1;0)A(N;0)
\nonumber \\
&&\:\:\:\: =
\frac{\sum_{i=1}^{N}k^{25}_{(i)}}{\sqrt{\tau}}
A(1,2,\cdots,N-2;1)A(N-1;0)A(N;0),
\end{eqnarray}
and the extra factor 
$\frac{\sum_{i=1}^{N}k^{25}_{(i)}}{\sqrt{\tau}}$ 
appears in front of the original term, again.
We can also show that all the terms in ${\cal A}^{(N,1)}$
get the same extra factor
through the modification and we can conclude that
\begin{eqnarray}
{\cal A}^{(N,2)}
=\frac{\sum_{i=1}^{N}k^{25}_{(i)}}{\sqrt{\tau}}
{\cal A}^{(N,1)}.
\end{eqnarray}
The same procedure holds for the calculation of 
${\cal A}^{(N,n)}$ from ${\cal A}^{(N,n-1)}$ and
therefore we can conclude that
\begin{eqnarray}
{\cal A}^{(N,n)}
=
\left(
\frac{\sum_{i=1}^{N}k^{25}_{(i)}}{\sqrt{\tau}}
\right)^{n}
{\cal A}^{(N,0)}.
\end{eqnarray}

\end{document}